\documentclass[a4paper,twocolumn,showpacs,prl]{revtex4}
\usepackage{mathrsfs}
\usepackage{amsfonts}
\usepackage{amsmath}
\usepackage{amssymb}
\usepackage{graphicx}
\usepackage{float}

\begin{document}

\title{Anomalous Electron Trajectory in Topological Insulators}
\author{Likun Shi$^{1}$, Shoucheng Zhang$^{2}$, Kai Chang$^{1}$}
\affiliation{$^{1}$SKLSM, Institute of Semiconductors, Chinese Academy of Sciences, P.O.
Box 912, Beijing 100083, China}
\affiliation{$^{2}$Department of Physics, Stanford University, Stanford, CA94305}

\begin{abstract}
We present a general theory about electron orbital motions in topological
insulators. An in-plane electric field drives spin-up and spin-down
electrons bending to opposite directions, the skipping orbital motions, a
counterpart of the integer quantum Hall effect, are formed near the boundary
of the sample. The accompanying Zitterbewegung can be found and controlled
by tuning external electric fields. Ultrafast flipping electron spin leads
to a quantum side jump in topological insulator, and a snake orbit motion in
two-dimensional electron gas with spin-orbit interactions. This feature
provides us a new way to control electron orbital motion by manipulating
electron spin.
\end{abstract}

\maketitle

The time-reversal invariant topological insulator (TI) is a new state of
quantum matter possessing insulating bulk and metallic edge or surface
states, which shows a linear massless Dirac dispersion {\cite%
{XLQi,KaneReview}}. TIs are distinguished from a normal band insulator by a
nontrivial topological invariant $Z_{2}$ characterizing its band structure.
The quantum spin Hall effect was proposed in graphene {\cite{Kane01}} and
HgTe quantum wells {\cite{SCZhang01} }. The existence of edge and surface
states was confirmed by the recent experiment in HgTe quantum wells {\cite%
{SCZhang02}} and the angle-resolved photoemission spectroscopy experiments {%
\cite{Hasan01,ZXShen}}. Due to its unique band structure, TI is a good
testbed for observing relativistic effects, predicted by the Dirac equation.
For instance, the Klein's paradox and Zitterbewegung (ZB).

So far, the most previous works in the rapid growing field of TI focused on
exploring new TIs and its transport and magnetic properties. Relatively,
electron dynamics in TIs is unexplored. In this work we show that the
quantum spin Hall effect and quantum anomalous Hall effect {\cite{SCZhang06}}
can be understood from anomalous electron orbital motions in TIs. These
anomalous electron orbital motions in topological insulators give naturally
a clear picture about the origin of the edge states, a counterpart of the
skipping orbital motion in the integer quantum Hall effect. By applying a
series of magnetic field pulses to flip electron spin quickly, a quantum
side-jump behavior and a snake-orbit motion can be found for electrons in
TIs and normal 2DEG with the spin-orbit interactions (SOIs), respectively.
The trembling motion, i.e., the ZB, can be controlled by changing an
in-plane electric field and the initial momentum of the electron wavepacket.

We consider the single-particle Hamiltonian of electron at low-energy regime
in the presence of a uniform electric field $\boldsymbol{E}$
\begin{align}
H & =H_{0}(k)+V(\boldsymbol{r}),  \notag \\
& ={\epsilon}(k)I+\sum_{i=1}^{3}d_{i}^{0}(k){\sigma}_{i}-e\boldsymbol{E}
\cdot\boldsymbol{r},
\end{align}
where $\sigma_{i}$($i=1,2,3$) is the Pauli matrix, ${\epsilon}(k)$ is the
kinetic energy. The different forms of the $d_{i}$ can be used to describe
the various important systems: 1) the two-dimensional electron gas with the
Rashba and Dresselhaus SOIs with $d_{1}^{0}=-\alpha k_{y}-\beta k_{x}$, $%
d_{2} ^{0}=\alpha k_{x}+\beta k_{y}$, $d_{3}^{0}=0$, where $\alpha$ and $%
\beta$ describe the strengths of the Rashba and Dresselhaus SOIs,
respectively {\cite{Loss}}; 2) single layer graphene with $%
d_{1}^{0}=v_{F}\hbar k_{x}$, $d_{2}^{0}=v_{F}\hbar k_{y}$, $d_{3}^{0}=0$ {%
\cite{Zawadzki01}}; 3) bilayer graphene with $d_{1}^{0}=-%
\hbar^{2}(k_{x}^{2}-k_{y}^{2})/2m$, $d_{2} ^{0}=-\hbar^{2}k_{x}k_{y}/m$, $%
d_{3}^{0}=0$ {\cite{Zawadzki01}}; 4) the two-dimensional TI HgTe quantum
wells with an inverted band structure with $d_{1}^{0}=Ak_{x}$, $%
d_{2}^{0}=Ak_{y}$, $d_{3}^{0}=M-Bk^{2}$ {\cite{SCZhang01} }; 5)
three-dimensional TIs, e.g., Bi$_{2}$Se$_{3}$ and Bi$_{2}$Te$_{3}$, whose
Hamiltonian would be extended to $H=\sum_{\mu,\nu=0}^{3}d_{\mu\nu}
^{0}(k)\sigma_{\mu}\otimes\sigma_{\nu}-e\boldsymbol{E}\cdot\boldsymbol{r}$,
in which $\sigma_{0}=I_{2\times2}$, and $d_{00}^{0}={\epsilon}(k)$, $d_{03}
^{0}=M(k)$, $d_{31}^{0}=A_{1}k_{z}$, $d_{11}^{0}=A_{2}k_{x}$, $d_{12}
^{0}=A_{2}k_{y}$ while other $d_{\mu\nu}^{0}$s are zero {\cite%
{Hasan02,SCZhang03}}.

In the absence of a uniform electric field, the electron position operator $%
y_{H}(t)$ evolving with the time $t$ can be obtained as
\begin{align}
y_{H}(t) & =e^{iHt/{\hbar}}ye^{-iHt/{\hbar}},  \notag \\
& =y(0)+(it/\hbar)[H,y]+{\frac{(it/\hbar)^{2}}{2!}}[H,[H,y]]+{\cdots },
\notag \\
& =y(0)+(it/\hbar){\epsilon}^{y}+{\sum\limits_{{n=1}}}{\frac{(it/\hbar)^{n}
}{n!}}T_{n},   \label{sum}
\end{align}
where ${\epsilon}^{y}=[{\epsilon},y]=-i{\partial}_{k_{y}}{\epsilon}$, $%
D^{0}=\sum_{j}d_{j}^{0}{\sigma}_{j}$, $T_{1}=[D^{0},y]=\sum_{j}d_{j} ^{y}{%
\sigma}_{j}=D^{y}$, $T_{2}=[D^{0},D^{y}]=\sum_{i,j}d_{i}^{0}d_{j} ^{y}[{%
\sigma}_{i},{\sigma}_{j}]$, $\cdots$, the $n$-th commutator $T_{n}$ $%
=[D^{0},T_{n-1}]$. Generally, the analytical expression of the above
summation is very difficult to obtain because the number of the commutators
increases hierarchically with increasing the order $n$.
\begin{figure}[ptb]
\centering
\includegraphics[scale=0.32]{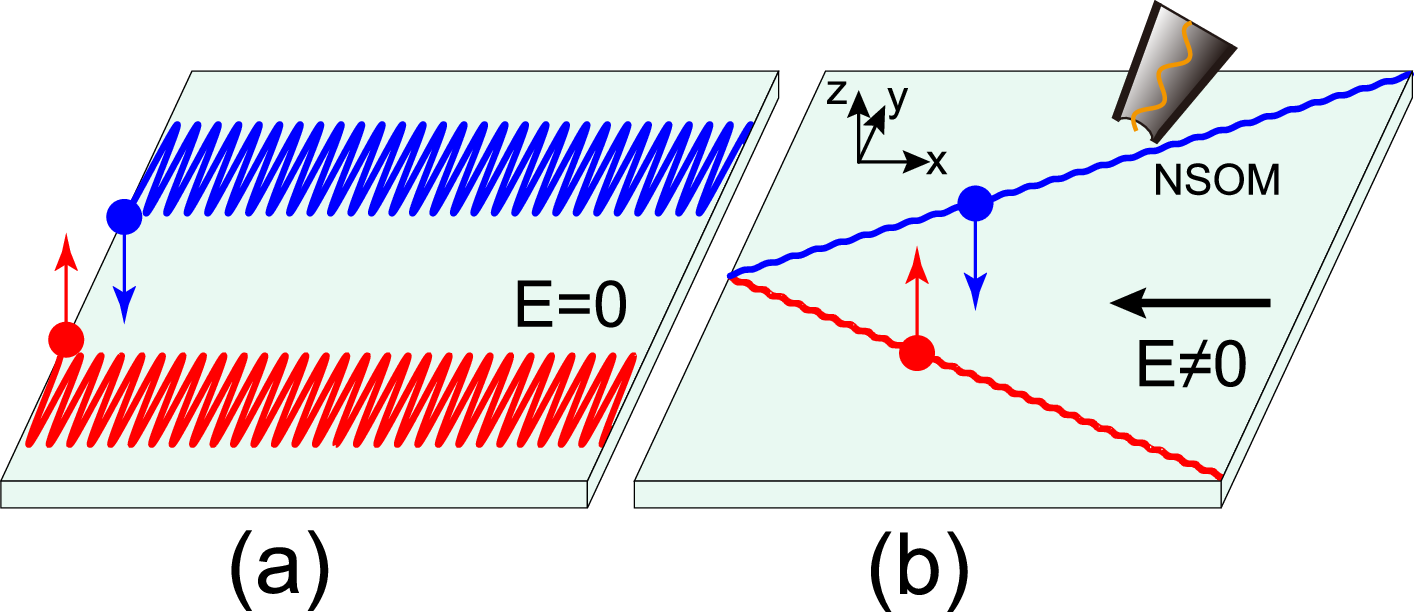}
\caption{(Color online) Schematic of electron orbital motion in a HgTe
quantum well with an inverted band without and with an in-plane driving
electric field [(a) and (b)]. The orbital motion can be detected by the
optical technique, e.g., the NSOM at the microwave frequency regime.}
\end{figure}

Interestingly, the equivalence between the commutation relationship $%
[\sigma_{i},\sigma_{j}]=2i\varepsilon_{ijk}\sigma_{k}$ and the vector cross
production ${(\boldsymbol{A}\times\boldsymbol{B})}_{k}=\varepsilon_{ijk}
A_{i}B_{j}\mathbf{e}_{k}$ (see the online supplementary material) offers us
a new way to solve this problem. In the absence of an external electric
field, the commutation between the spin operators is converted into the
vector product between them. We can transfer the algebra summation in Eq. (%
\ref{sum}) into the summations of the vector series $T_{2n+1}(T_{2n})(n=1,2,%
\cdots)$ utilizing the equivalence between the commutator $[D^{0},T_{n}]$
and the vector product ${D}^{0}\times{T}_{n}$, $T_{n}=[D_{0},T_{n-1}]\sim2i({%
D} ^{0}\times{T}_{n-1})$ (see the online supplementary material). One can
see that the vectors $T_{2n}$ and $T_{2n+1}$ ($n=0,1,2,...$) point along the
orthogonal directions while the lengths of the vectors ${T}_{n}$ varies with
the increasing of the order $n$ as $T_{2n+1}=-|2{D}^{0}|^{2}T_{2n-1}$, where
$|D_{0}|=(\sum_{i=1} ^{3}(d_{i}^{0})^{2})^{1/2}$. This character allows us
to get the analytical expression for the electron position operator
\begin{align}
y_{H}(t) & =y(0)+\frac{it}{\hbar}\left( {\epsilon}^{y}+T_{1}-\frac{T_{3} }{|{%
D}^{0}|^{2}}\right)  \notag \\
& +\frac{T_{2}}{2|{D}^{0}|^{2}}\left[ \cos\left( 2|{D}^{0}|t/\hbar\right) -1%
\right]  \notag \\
& +\frac{iT_{3}}{2|{D}^{0}|^{3}}\left[ \sin\left( 2|{D}^{0}|t/\hbar\right) %
\right] .
\end{align}

This analytical expression consists of the initial position (the first
term), the propagating term (the second term) and the \textit{Zitterbewegung}
term (the last two terms). This ZB term describes the rapid trembling motion
which has no the classical correspondences. ZB, a novel relativistic quantum
orbital motion, is an inherent hallmark of the Dirac equation, first
predicted by Schr\"{o}dinger in 1930 {\cite{Schrodinger}}. This rapid
quivering motion of electron arises from the superposition between the
positive- and negative- energy part of the spinor states and could be a
possible origin of electron spin {\cite{KHuang}}. Although there are many
proposals about observation of the ZB in a two-dimensional electron gas
(2DEG) with the SOIs {\cite{Loss,Schliemann}, graphene \cite{Zawadzki01}}
and a Cooper pair in superconductors {\cite{Cannata,Lurie}} and related
theoretical analysis {\cite{refreeC}}, but only observed experimentally in
the trapped ion {\cite{Solano1,Solano2,Roos}} and cold atom {\cite{Clark}}
systems. However, this prediction has still never been observed for
electrons in free space, this is because the ZB displays an extremely high
frequency $\omega_{Z}=2m_{0}c^{2}/\hbar\approx1.5\times10^{21}$ Hz and a
tiny amplitude $\lambda_{Z}=\hbar/m_{0}c\approx3.9\times10^{-4}$ nm ($m_{0}$
is the electron mass). The detection of the oscillation with such high
frequency and negligibly small amplitude is beyond the reach of the
present-day experimental technique.

It is natural to imagine that the frequency of the ZB should be decreased
when the energy gap decreases. TI could be a good testbed to observe the ZB
due to its narrow bulk gap which ranges from 10meV to 0.3 eV. The expression
of ZB (see Eq. [3]) agrees exactly with the previous theoretical works for
various systems. For a 2DEG with the Rashba and Dresselhaus SOIs {\cite{Loss}%
}, ${\epsilon}^{y}+T_{1}=-i{\hbar}^{2}{k}_{x}/m-i(\alpha{\sigma}_{y}-\beta {%
\sigma}_{x})$, $T_{2}=(\alpha^{2}-\beta^{2})k_{y}\sigma_{z}$, $T_{3}
=(\alpha^{2}-\beta^{2})(k_{x}\sigma_{y}-k_{y}\sigma_{x})\Sigma$, where $%
\Sigma=\alpha(k_{x}\sigma_{x}+k_{y}\sigma_{y})+\beta(k_{x}\sigma_{y}
+k_{y}\sigma_{x})$, $|D^{0}|=(\alpha^{2}+\beta^{2})k^{2}+4\alpha\beta{k_{x} }%
{k_{y}}$. For a single layer graphene {\cite{Zawadzki01}}, ${\epsilon}
^{y}+T_{1}=-i{\hbar}{v_{F}}\sigma_{x}$, $T_{2}={\hbar}^{2}{v_{F}}^{2}{k_{y} }%
{\sigma_{z}}$, $T_{3}={\hbar}^{3}{v_{F}}^{3}{k_{y}}({k_{x}}{\sigma_{y} -{%
k_{y}}{\sigma_{x}}})$, $|D^{0}|={v_{F}{\hbar}k}$. The two-dimensional TI
HgTe quantum wells with an inverted band structure {\cite{SCZhang01}}, ${%
\epsilon}^{y}+T_{1}=-i(2Dk_{y}+A\sigma_{y})$, $T_{2}=-2A(M-Bk^{2})\sigma
_{x}+2A^{2}k_{x}\sigma_{z}$, $T_{3}=4i\{A^{3}k_{x}k_{y}\sigma_{x}
-[A(M-Bk^{2})^{2}+A^{3}k_{x}^{2}]\sigma_{y}+A^{2}(M-Bk^{2})k_{y}\sigma_{z}\}$
(see Fig. 1(a)). The second term ${\epsilon}^{y}+T_{1}-T_{3}/|{D}^{0}|^{2}$
represents the classical uniform rectilinear motion, the third and the
fourth terms $T_{2}$ and $T_{3}$ describe the ZB with the oscillating
frequency $2|{D}^{0}|/\hbar$. The trembling frequency is determined by the
strength of the spin-orbit interaction of the interband coupling $d_{i}^{0}$%
. For all above examples, these agreements demonstrate the validity of the
diagram technique developed by us.

In two-terminal transport experiments, an external voltage is applied
between the source and drain, generating an in-plane electric field. In the
presence of a uniform in-plane electric field, it is difficult to get the
analytical expression of the electron position operator as discussed above.
Instead, we can calculate the electron position $\mathbf{r}(t)$ based on the
equation of motion $i\hbar\dot{\mathbf{r}}=[\mathbf{r},H]$. We consider a
two-dimensional topological insulator, a HgTe quantum well with an inverted
band structure described by the BHZ model {\cite{SCZhang01}}. The
single-particle effective Hamiltonian for the electron is
\begin{equation}
H(k)_{\uparrow\downarrow}={H}_{0}(k)_{\uparrow\downarrow}+V(x_{i}),
\end{equation}
where ${H}_{0}(k)_{\uparrow\downarrow}=C-Dk^{2}\pm Ak_{x}\sigma^{x}
+Ak_{y}\sigma^{y}+(M-Bk^{2})\sigma^{z}$, $V(\mathbf{x})=-e\mathbf{E\cdot r}%
\sigma^{z}$. $A$, $B$, $C$, $D$, and $M$ are the parameters determined by
the thickness of the quantum well {\cite{MKonig}.} Consider an electron
injected in the Gaussian wave packet $\varphi(r)=\frac{1}{2\pi}\frac{{d} }{%
\sqrt{\pi}}\int d^{2}ke^{-d^{2}(\overrightarrow{k}-\overrightarrow{k}
_{0})^{2}/2}e^{i\overrightarrow{k}\overrightarrow{r}}\left\vert \uparrow
\right\rangle $ with the spatial width $(\Delta x)^{2}=(\Delta y)^{2}=d^{2}/2
$ and spin pointing along the $z$ axis, perpendicular to the quantum well
plane. The guiding center of the wave packet ${\langle}y_{H}(t){\rangle}$
can be calculated numerically by the Heisenberg equation of motion $d{\langle%
} y_{H}(t){\rangle/dt=(ih)}^{-1}\left\langle \left[ y_{H}(t),H\right]
\right\rangle $. From the numerical results (the red and blue curves in Fig.
1(c)), the spin-up and spin-down electron driven by an in-plane electric
field $E_{x}$ bend to the opposite direction along the $y$ axis accompanying
with a trembling behavior, i. e., the ZB. Notice that the electron energy $%
\left\langle E\right\rangle =C+M-(B+D)/d^{2}-(B+D)k_{0}^{2}$ locates in the
bulk gap, which is the important difference from the spin Hall effect in a
conventional 2DEG and/or metal with the SOIs.

In order to understand this surprising feature, we analyze the equation of
motion utilizing the adiabatic approximation, i.e., separating the fast
trembling motion and slow orbital motion $y_{\uparrow\downarrow}
=y_{\uparrow\downarrow}^{\text{orb}}+y_{\uparrow\downarrow}^{\text{ZB}}$ {%
\cite{SCZhang04,SCZhang05}}. First we perform a unitary transformation $U(k)$
to diagonalize the Hamiltonian ${H}_{0}(k)$, i. e., $\tilde{H}_{0}
=U(k)_{\uparrow\downarrow}H_{0}(k)_{\uparrow\downarrow}U^{\dag}(k)_{\uparrow
\downarrow}$. The potential term becomes $V(\tilde{D}_{i})$, where the
covariant derivative $\tilde{D}_{i}=i\partial_{k_{i}}-\tilde{A}_{i}$, and $%
\tilde{A} _{i}(k)_{\uparrow\downarrow}=-i\cdot U(k)_{\uparrow\downarrow
}\partial_{k_{i} }U^{\dag}(k)_{\uparrow\downarrow}\cdot U(k)_{\uparrow
\downarrow}\sigma^{z}U^{\dag}(k)_{\uparrow\downarrow}$ behaves like a gauge
field. Adopting the adiabatic approximation, i.e., neglecting the
off-diagonal matrix elements of $\tilde{A}_{i}$, the resulting gauge field $%
A_{i}$ only contains a diagonal matrix which gives a non-zero associated
field strength $F_{ij}=i[D_{i},D_{j}]$, where $x_{i}\rightarrow
D_{i}=i\partial_{k_{i}} -A_{i}$, then the effective Hamiltonian becomes
\begin{equation}
H_{\uparrow\downarrow}^{\text{eff}}=\tilde{H}_{0}(k)-e\sum\limits_{i=x,y}
E_{i}D_{i,\uparrow\downarrow},
\end{equation}
where $\tilde{H}_{0}(k)=C-Dk^{2}-\sqrt{A^{2}k^{2}+(M-Bk^{2})^{2}}\sigma^{z}$
for both spin-up (down) states. The nontrivial property of the Hamiltonian
is revealed through the nontrivial commutation relations $[k_{i},k_{j}]=0$, $%
[D_{i},k_{j}]=i\delta_{ij}$, $[D_{i},D_{j}]=-iF_{ij}$, the effective
\textquotedblleft Lorentz forces" in momentum space felted by spin-up or
spin-down electron is
\begin{equation*}
F_{xy}(k)_{\uparrow\downarrow}=\pm\frac{A^{2}(M^{2}-B^{2}k^{4})}{2[A^{2}
k^{2}+(M-Bk^{2})^{2}]^{2}}.
\end{equation*}
The equation of motion for the spin-up/down electron can be written as
\begin{align}
{\dot{x}}_{\uparrow\downarrow} & =u_{x}\pm\left\vert F_{xy}\right\vert {\dot{%
k}}_{y},  \notag \\
{\dot{y}}_{\uparrow\downarrow} & =u_{y}\mp\left\vert F_{xy}\right\vert {\dot{%
k}}_{x}, \\
k_{i} & =k_{i0}+\lambda eE_{i}t/\hbar,  \notag
\end{align}
where $u_{i}$ represents the usual group velocity $\partial\tilde{H}
_{0}/\partial k_{i}$, and $\lambda=\pm1$ stands for the negative (positive)
branch of the energy spectrum. When $k_{y0}=0$, the orbital trajectory of
spin-up or spin-down electron at small $k$ is (see the online supplementary
material)
\begin{equation}
y_{\uparrow\downarrow}^{\text{orb}}=\mp\frac{A^{2}}{2M^{2}}k_{x}+O(k)^{3}.
\end{equation}

Clearly, one can see that the spin-up and spin-down electrons feel an
opposite force $F_{xy}(k)_{\uparrow\downarrow}$, respectively, which pushes
them against the opposite direction. The trajectories obtained from Eq. [7],
which neglects the fast trembling motion by the adiabatic approximation. The
analytical expression for the trembling motion, i.e., the ZB, at small $k$
can also be obtained (see the online supplementary material)
\begin{equation}
y_{\uparrow\downarrow}^{\text{ZB}}=\pm\frac{A^{2}}{2M^{2}}\frac{eE_{x}} {%
\Delta\epsilon(t)}\sin(\frac{\Delta\epsilon(t)}{\hbar}t),
\end{equation}
where $\Delta\epsilon(t)=2\sqrt{A^{2}k^{2}+(M-Bk^{2})^{2}}=2|M|+O(k^{2})$.
At small in-plane momentum $k$, the kinetic energy can be neglected, the
frequency of amplitude of the ZB are both determined by $M$, the gap of HgTe
QW which can be tuned by the thickness of the QW. The total electron
trajectories $y_{\uparrow\downarrow}=y_{\uparrow\downarrow}^{\text{orb}
}+y_{\uparrow\downarrow}^{\text{ZB}}$ obtained from Eq. [7] and [8] agree
well with the numerical results for spin-up and spin-down electron incident
cases (see Fig. 1(c)). The inset in Fig. 1(c) shows the amplitude of the ZB
increases rapidly with decreasing the gap $|M|$ of HgTe QW with an inverted
band, and a linear dependence on the strength of the in-plane electric field
$E_{x}$. For example, the amplitude of the ZB is about $21.2\mathring{A}$
when $M=2.5$meV and $E_{x}=10$V cm$^{-1}$, which are within the reach of the
state-of-art experimental techniques. By adjusting the electric fields $E_{x}
$ and the bulk gap of HgTe QWs, one can tune the amplitude of the ZB
significantly (see Fig. 2(a)), and make it possible to observe it in TIs.
\begin{figure}[ptb]
\centering
\includegraphics[scale=0.36]{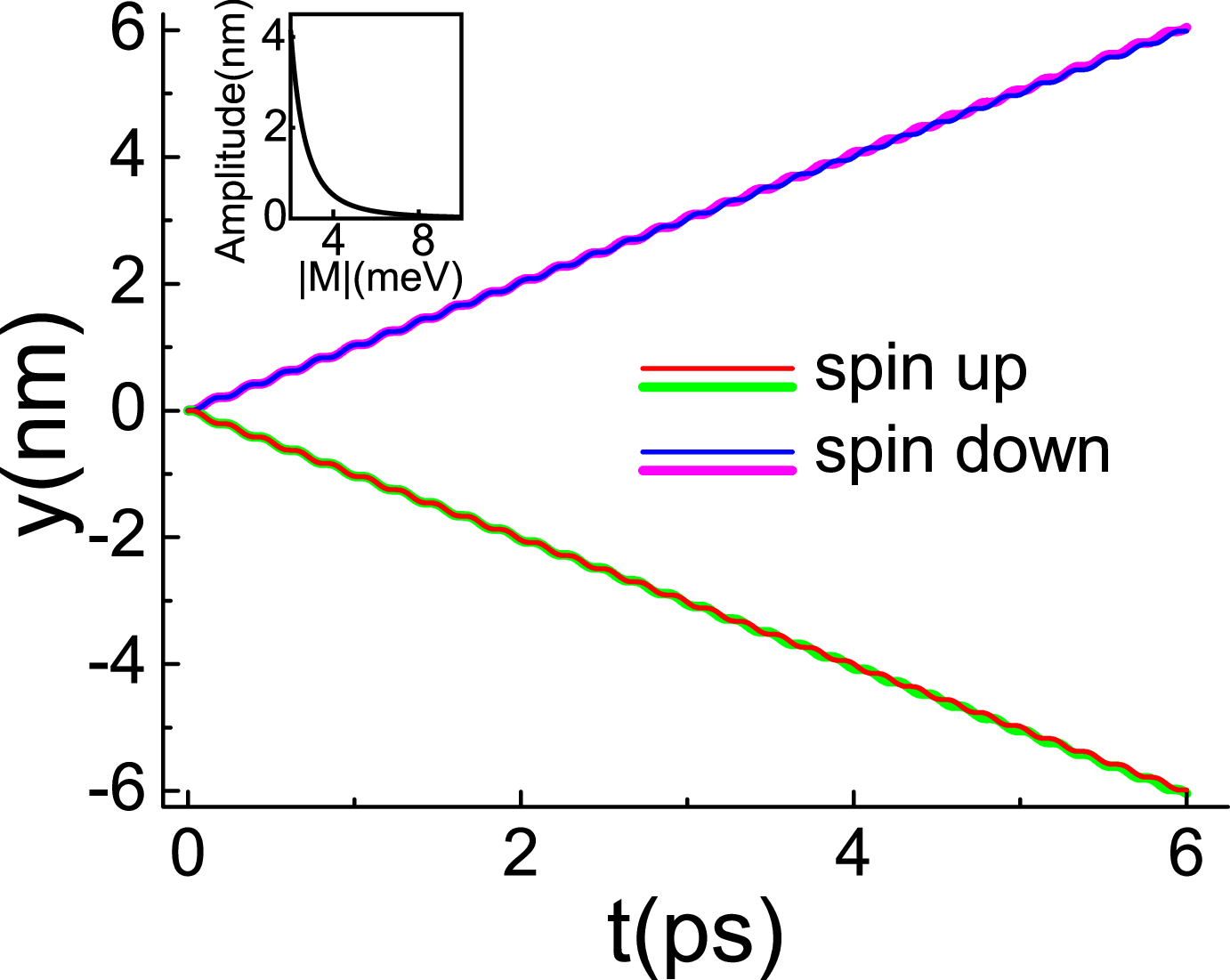}
\caption{ (color online). The trajectories of spin-up (the red and green
curves) and spin-down (the blue and purple curves) electrons incident with
the initial momenta $k_{x0}=0.001$nm$^{-1}$, the electric field $E_{x}=10$%
V/cm. The red and blue (green and purple) curves denote the numerical
(analytical) results. The parameters used in the calculation are adopted
from {\protect\cite{MKonig}}. The inset shows that the amplitude of the
transverse ZB $y^{\text{ZB}}$ as a function of the bulk gap of HgTe QW $|M|$.
}
\end{figure}

For the realistic experimental sample, we need to consider the boundary
effect, for instance, a spin Hall bar with a finite width. The spin-up and
spin-down electrons will bend to opposite edges. Fig. 3 (b) shows clearly
that spin-up and spin-down edge states appear near the opposite boundaries
of the sample (the shaded regions). From the inset, one can see the skipping
orbital motion of electrons in the edge states, since electrons bending to
the edges will be bounced back when it hits the hard-wall boundary {\cite%
{HardWall}, and pushed to the boundary again by the driving force (or the
effective "Lorentz" force) $F_{xy}(k)_{\uparrow\downarrow}$. This is a
counterpart of the skipping orbital motion in the integer quantum Hall
effect, but spin-up and spin-down electrons feel opposite the effective
"Lorentz" forces. While in a normal 2DEG with the SOIs, spin-up and
spin-down electrons show a ballistic side jump in opposite directions which
was already found before {\cite{Schliemann}, but no edge states can be found
near the sample boundary. Instead, electrons will oscillate between the
opposite boundaries (see Figs. 3(c) and 3(d)). } }

Similarly, the quantum anomalous spin Hall effect might also be understood
from the electron dynamics in TIs. Considering a ferromagnetic TI, a giant
Zeeman splitting results in a spin-up inverted and a spin-down normal band
structure, we assume these magnetic ions polarized along $z$ axis. This
situation corresponds to the BHZ Hamiltonian with the positive and negative $%
M$ for spin-up and spin-down electrons, respectively. From the above
discussions, the spin-up (down) electrons show normal oscillating (bending)
trajectory (see Figs. 3). Therefore, the edge states only appear for
spin-down electrons, i.e., the quantum anomalous spin Hall effect. Similar
behavior can also be found in three-dimensional topological insulators,
because the form of the Hamiltonian of 3D TI is almost the same as that of
2D TI. Therefore one can expect that the bending trajectory of electrons in
3D TIs will lead to the topological surface states.
\begin{figure}[ptb]
\centering
\includegraphics[scale=0.36]{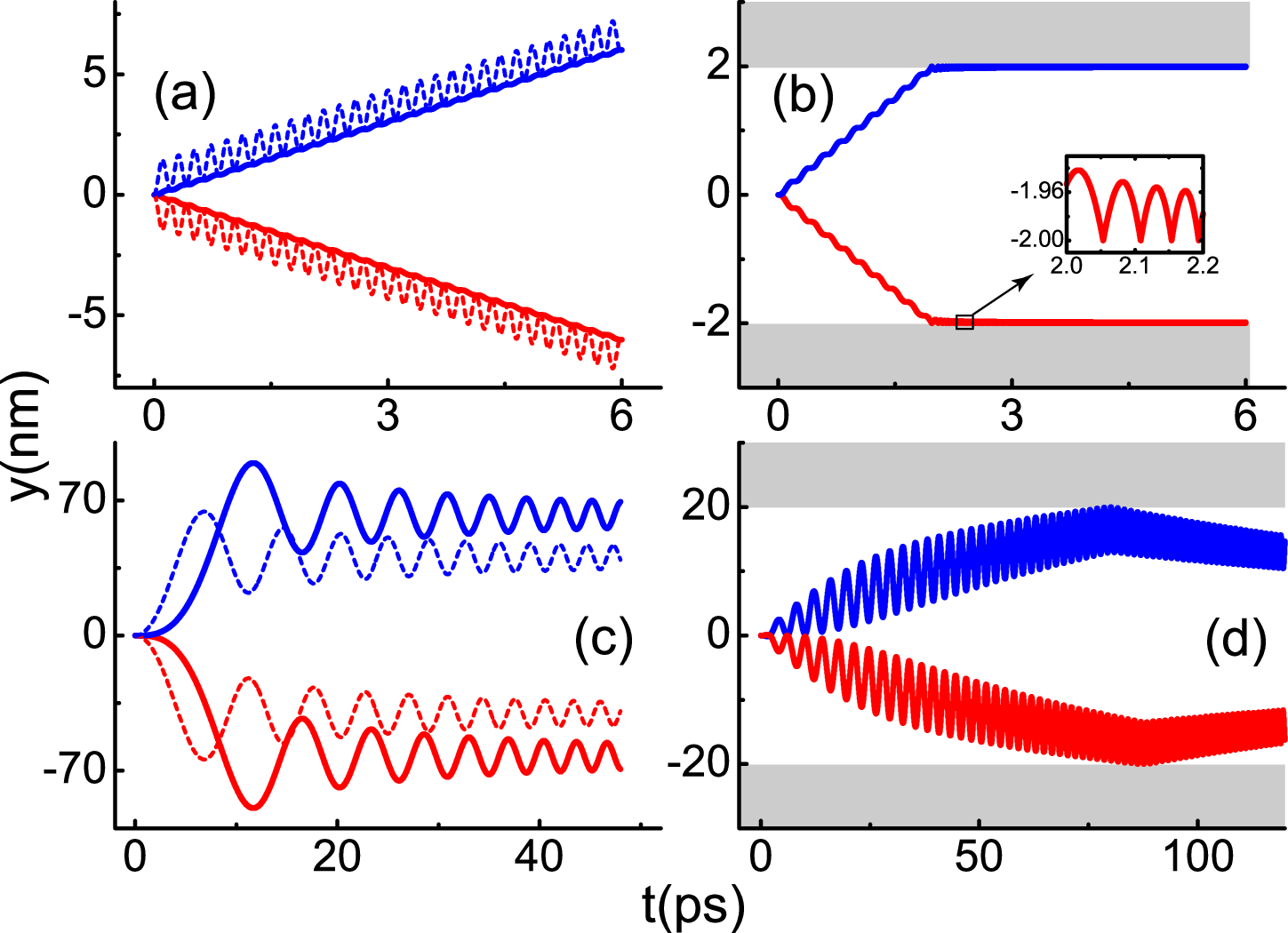}
\caption{ (color online). The trajectories of spin-up (red) and spin-down
(blue) incident electrons under a uniform electric field. The in-plane
electric field $E=10$V cm$^{-1}$. The electron injected into HgTe quantum
well ($M=-10$ meV) (a) has a initial incident momentum $k_{x0}=0$ for the
solid curves and $k_{x0}=0.001$nm$^{-1}$ for the dashed curves; (b) The same
as (a), but with a hard-wall boundary (the shaded regions). The inset
amplifies the trajectory near the boundary, showing the skipping orbital
motion. (c) and (d) The same as (a) and (b), but in a conventional GaAs 2DEG
with the Rashba spin-orbit interaction ($\protect\alpha=10$ meV nm taken
from Ref. [4]). The solid and dashed curves corresponds to $k_{x0}=0$ and $%
k_{x0}=0.01$nm$^{-1}$ , respectively.}
\end{figure}

In a spin-orbit system, usually electron spin can be manipulated by control
its orbital motion, such as spin transistor {\cite{Spintransistor}}. And
\textit{vise versa}, it is also possible to control electron orbital motion
by manipulating electron spin. In a 2D TI, e.g., a 2DEG in a HgTe QW with an
inverted band structure, electron spin flipping leads to an interesting
quantum side jump (see Fig. 4(a)), which is the manifestation of the
spin-dependent ZB and caused by the spin-dependent driving force. While in
the 2DEG with the SOIs, an interesting snake orbit motion can be found (see
Fig. 4(b)). This ultrafast spin flipping process can be achieved by applying
a series of specifically shaped magnetic field pulses (see the insets in
Figs. 4(c) and 4(d)), which was used to flip the magnetization of the
magnetic random access memory devices {\cite{pulseB}}. The width of the
magnetic feild pulses is determined by the magnitudes of the magnetic field
pulses $B_{0}$, usually it will take longer (shorter) time to flip electron
spins for weak (strong) magnetic field $B_{0}$.
\begin{figure}[tbp]
\centering
\includegraphics[scale=0.36]{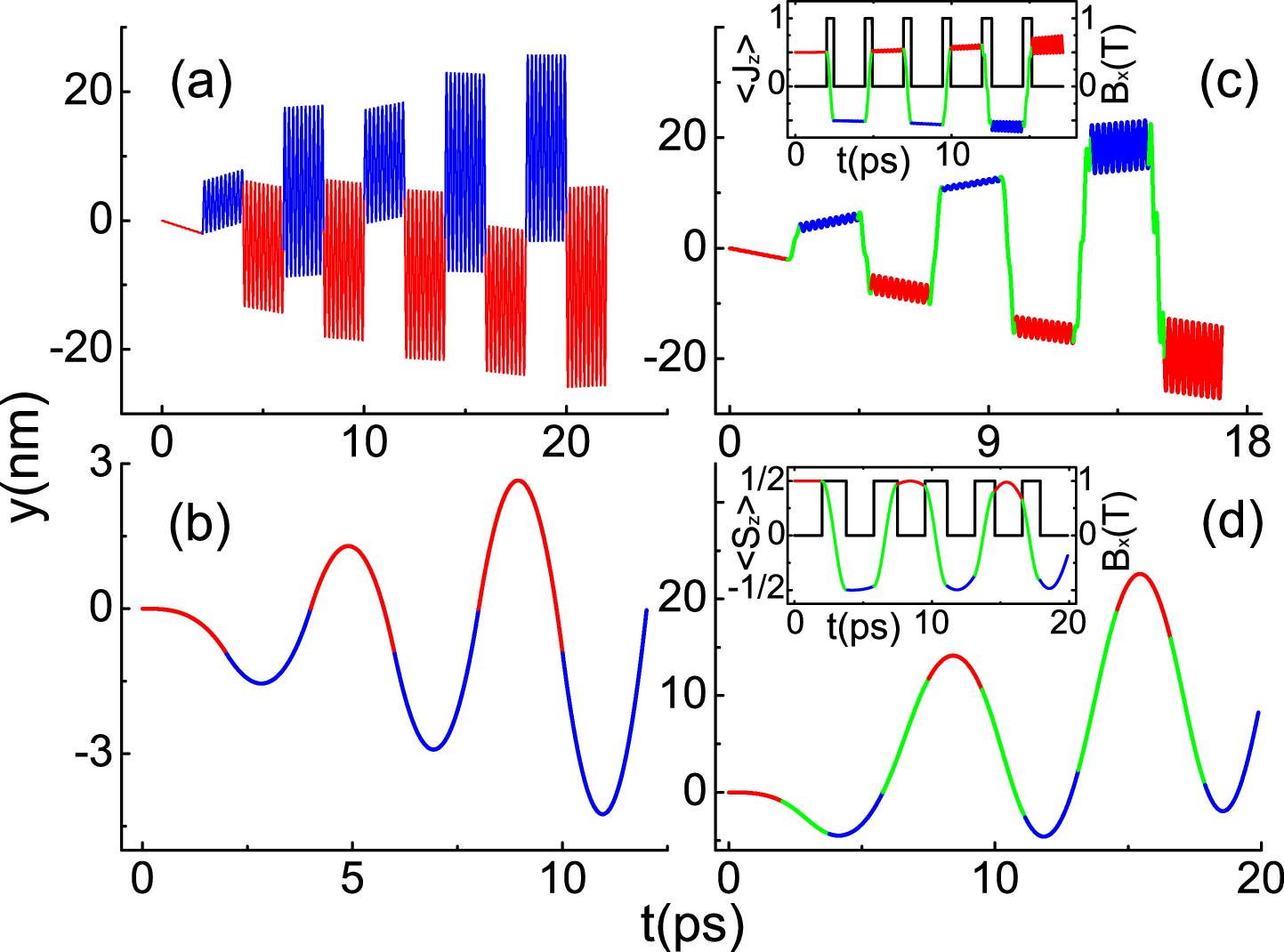}
\caption{ (color online). The trajectories of a initial spin-up electron
driven by a uniform electric field in HgTe QW (a) and a GaAs 2DEG with the
SOIs (b). The red and blue curves denote the spin-up and spin-down
electrons, respectively. The electron spin is flipped periodically at a time
interval $t=2$ps. (c) and (d), the same as (a) and (b), but for a realistic
case: a series of the square- shaped magnetic field pulses with the
magnitude $B_{0}=1$T. The parameters are the same with those in Fig. 3, $%
k_{0}=0$.}
\end{figure}

Finally, we comment on how to detect these interesting orbital motions
experimentally. We propose a near-field scanning optical microscope (NSOM)
technique in microwave or THz regime to detect the edge states, ZB, quantum
side jump and snake-orbit motions. The NSOM technique breaks the far field
resolution limit by exploiting the properties of evanescent waves. With this
technique, the resolution of the image is limited by the size of the
detector aperture and not by the wavelength of the illuminating light. A
spatial lateral resolution of the NSOM can approach to 20 nm {\cite{NSOM}}.
A microwave light beam is applied at a specific spatial position through a
NSOM (see Fig. 1(b)), the absorption of the microwave beam changes when
electrons pass below. This experimental technique was used to detect the
topological edge states in a HgTe QW very recently.

In summary, we give an intuitive physical picture about the dynamical origin
of the edge states in TIs, a skipping orbit motion. By flipping electron
spin quickly, one can control the electron orbital motion efficiently. An
interesting quantum side jump and snake orbital motion can be found in the
TIs and 2DEGs with the SOIs. This feature provides us a new way to control
electron orbital motion by manipulating electron spin.

\begin{acknowledgments}
This work was supported by the NSFC Grants No. 10934007 and the grant No.
2011CB922204 from the MOST of China, and S.C.Z. is supported by the DARPA
Program on Topological Insulators.
\end{acknowledgments}


\begin{thebibliography}{99}
\bibitem{XLQi} X. L. Qi and S. C. Zhang, Phys. Today \textbf{63}, 33 (2010).

\bibitem{KaneReview} M. Z. Hasan and C. L. Kane, Rev. Mod. Phys. \textbf{82}%
, 3045 (2010).

\bibitem{Kane01} C. L. Kane, E. J. Mele, Phys. Rev. Lett. \textbf{95},
146802 (2005).

\bibitem{SCZhang01} B. A. Bernevig, T. L. Hughes, S. C. Zhang, Science
\textbf{314}, 1757 (2006).

\bibitem{SCZhang02} M. K\"{o}nig, S. Wiedmann, C. Br\"{u}ne, A. Roth, H.
Buhmann, L. W. Molenkamp, X. L. Qi, S. C. Zhang, Science \textbf{318}, 766
(2007).

\bibitem{Hasan01} D. Hsieh, D. Qian, L. Wray, Y. Xia, Y. S. Hor, R. J. Cava,
M. Z. Hasan, Nature (London) \textbf{452}, 970 (2008).

\bibitem{ZXShen} Y. L. Chen, J. G. Analytis, J. H. Chu, Z. K. Liu, S. K. Mo,
X. L. Qi, H. J. Zhang, D. H. Lu, X. Dai, Z. Fang, S. C. Zhang, I. R. Fisher,
Z. Hussain, Z. X. Shen, Science \textbf{352}, 178 (2009).

\bibitem{SCZhang06} C. X. Liu, X. L. Qi, Xi Dai, Z. Fang and S. C. Zhang,
Phys. Rev. Lett. \textbf{101}, 146802 (2008).

\bibitem{Loss} J. Schliemann, D. Loss, and R. M. Westervelt, Phys. Rev.
Lett. \textbf{94}, 206801 (2005).

\bibitem{Zawadzki01} T. M. Rusin, W. Zawadzki, Phys. Rev. B \textbf{76},
195439 (2007).

\bibitem{Hasan02} Y. Xia, D. Qian, D. Hsieh, L. Wray, A. Pal, H. Lin, A.
Bansil, D. Grauer, Y. S. Hor, R. J. Cava and M. Z. Hasan, Nature Phys.
\textbf{5}, 438 (2009).

\bibitem{SCZhang03} H. J. Zhang, C. X. Liu, X. L. Qi, X. Dai, Z. Fang and S.
C. Zhang, Nature Phys. \textbf{5}, 438 (2009).

\bibitem{Schrodinger} E. Shr\"{o}dinger, Sitzungsb. Preuss. Akad. Wiss.
Phys.-Math. K1. \textbf{24}, 418 (1930).

\bibitem{KHuang} K. Huang, Am. J. Phys. \textbf{20}, 479 (1952).

\bibitem{Schliemann} J. Schliemann, Phys. Rev. B. \textbf{75}, 045304 (2007).

\bibitem{Cannata} F. Cannata, L. Ferrari, and G. Russo, Solid State Commun.
\textbf{74}, 309 (1990); L. Ferrari and G. Russo, Phys. Rev. B \textbf{42},
7454 (1990); F. Cannata and L. Ferrari, Phys. Rev. B \textbf{44}, 8599
(1991).

\bibitem{Lurie} D. Luri\'{e} and S. Cremer, Physica (Amsterdam) \textbf{50},
224 (1970).

\bibitem{refreeC} R. Winkler, U. Z\"{u}licke and J. Bolte, Phys. Rev. B.
\textbf{75}, 205314 (2007); J. Cserti, G. D\'{a}vid, Phys. Rev. B. \textbf{82%
}, 201405 (2010); W. Zawadzki, T. M. Rusin, J. Phys. Condens. Matter \textbf{%
23}, 143201 (2011).

\bibitem{Solano1} L. Lamata, J. Le\'{o}n, T. Sch\"{a}tz, and E. Solano,
Phys. Rev. Lett. \textbf{98}, 253005 (2007).

\bibitem{Solano2} A. Bermudez, M. A. Martin-Delgado, and E. Solano, Phys.
Rev. A. \textbf{76}, 041801(R) (2007).

\bibitem{Roos} R. Gerritsma, G. Kirchmair, F. Z\"{a}hringer, E. Solano, R.
Blatt, C. F. Roos, Nature (London) \textbf{463}, 68 (2010).

\bibitem{Clark} J. Y. Vaishnav, C. W. Clark, Phys. Rev. Lett. \textbf{100},
153002 (2008).

\bibitem{MKonig} M. K\"{o}nig, H. Buhmann, L. W. Molenkamp, T. Hughes, C. X.
Liu, X. L. Qi, S. C. Zhang, J. Phys. Soc. Jpn. \textbf{77}, 031007 (2008).

\bibitem{SCZhang04} S. Murakami, N. Nagaosa, S. C. Zhang, Science \textbf{301%
}, 1348 (2003).

\bibitem{SCZhang05} Z. F. Jiang, R. D. Li, S. C. Zhang, and W. M. Liu, Phys.
Rev. B. \textbf{72}, 045201 (2005).

\bibitem{HardWall} L. Fu and C. L. Kane, Phys. Rev. Lett. \textbf{100},
096407 (2008); B. Zhou, H. Z. Lu, R. L. Chu, S. Q. Shen, and Q. Niu, Phys.
Rev. Lett. \textbf{101}, 246807 (2008); E. G. Novik, P. Recher, E. M.
Hankiewicz, and B. Trauzettel, Phys. Rev. B. \textbf{81}, 241303 (2010) K.
Chang and W. K. Lou, Phys. Rev. Lett. \textbf{106}, 206802 (2011).

\bibitem{Spintransistor} S. Datta and B. Das, Appl. Phys. Lett. \textbf{56},
665 (1990); H. C. Koo, J. H. Kwon, J. Eom, J. Chang, S. H. Han, and M.
Johnson, Science \textbf{325}, 1515 (2009).

\bibitem{pulseB} Th. Gerrits, H. A. M. van den Berg, J. Hohlfeld, L. B\"{a}%
r, and Th. Rasing, Nature (London), \textbf{418}, 509 (2002); M. R. Freeman,
R. R. Ruf, and R. J. Gambino, IEEE TRANS. MAG. \textbf{27}, 4840 (1991).

\bibitem{NSOM} Y. Oshikane T. Kataoka, M. Okuda, S. Hara, H. Inoue, M.
Nakano, Sci. Technol. Adv. Mater. \textbf{8}, 181 (2007).
\end{thebibliography}
\end{document}


\title{Online Supplemental Material:
Anomalous Electron Trajectory in Topological Insulators}

\author{Likun Shi$^{1}$, Shoucheng Zhang$^{2}$, Kai Chang$^{1}$}
\affiliation{$^{1}$SKLSM, Institute of Semiconductors, Chinese Academy of Sciences, P.O.
Box 912, Beijing 100083, China}
\affiliation{$^{2}$Department of Physics, Stanford University, Stanford, CA94305}

\maketitle

\section{I. Diagram technique in the absence of external field}

We consider a single-particle Hamiltonian of electron in a general form
\begin{align}
H  &  =\epsilon\left(  k\right)  +\sum_{i}d_{i}\left(  k\right)  \sigma^{i}\\
&  =\epsilon\left(  k\right)  +D\left(  k\right) \nonumber
\end{align}
where $\epsilon\left(  k\right)  $ denotes the kinetic energy and the second
term describes the spin-orbit interactions or interband coupling.

In the Heisenberg picture, the position operator of electron $y_{H}(t)$
evolving with time $t$ can be obtained by
\begin{align}
y_{H}(t)  &  =e^{iHt/{\hbar}}ye^{-iHt/{\hbar}},\\
&  =y(0)+(it/\hbar)[H,y]+{\frac{(it/\hbar)^{2}}{2!}}[H,[H,y]]+{\cdots
},\nonumber\\
&  =y(0)+(it/\hbar){\epsilon}^{y}+{\sum\limits_{{n=1}}}{\frac{(it/\hbar)^{n}
}{n!}}T_{n},\nonumber
\end{align}
where we show the notions explicitly:
\begin{align}
{\epsilon}^{y}  &  =[{\epsilon},y]=-i{\partial}_{k_{y}}{\epsilon},\\
T_{1}  &  =[H,y]=[D,y],\nonumber\\
T_{2}  &  =[H,T_{1}]=[D,T_{1}],\nonumber
\end{align}
\[
{\vdots}
\]%
\[
T_{n}=[D,T_{n-1}].
\]

The above equation involves the infinite summation of the commutators $T_{n}$
and therefore it is not easy to obtain the analytical expression of the
position operator $y_{H}(t)$. In order to calculate the commutation in
arbitrary order and the summation, we develop a diagram technique as follow
which makes the lengthy calculation of the commutation much easier. Utilizing
the similarity between the commutation relationship $[{\sigma}_{i},{\sigma
}_{j}]=2i\varepsilon_{ijk}{\sigma}_{k}$ and the vector cross production
$\boldsymbol{A}\times\boldsymbol{B}=\varepsilon_{ijk}A_{i}B_{j}\boldsymbol{e}%
_{k}$, we obtain the following relationship
\begin{align}
D\left(  k\right)  =d_{1}{\sigma}^{1}+d_{2}{\sigma}^{2}+d_{3}{\sigma}^{3}  &
\rightarrow\boldsymbol{D}=\left(  d_{1},d_{2},d_{3}\right)  ,\\
{T}_{1}=[H,{y]}=t_{0}+t_{1}{\sigma}^{1}+t_{2}{\sigma}^{2}+t_{3}{\sigma}^{3}
&  \rightarrow\boldsymbol{T}_{1}=\left(  t_{1},t_{2},t_{3}\right)
,\nonumber\\
{T}_{2}=[H,{T}_{1}{]}=[D,{T}_{1}]  &  \rightarrow2i\left(  \boldsymbol{D}%
\times\boldsymbol{T}_{1}\right)  ,\nonumber\\
&  \cdots\nonumber\\
{T}_{n+1}=[H,{T}_{n}{]}=[D,{T}_{n}]  &  \rightarrow2i\left(  \boldsymbol{D}%
\times\boldsymbol{T}_{n}\right)  .\nonumber
\end{align}

These relationships can be illustrated in Fig. 1. From this figure, one can
see that the vectors $\boldsymbol{T}_{2n+1}$ and $\boldsymbol{T}_{2n}$
$(n=1,2,3\cdots)$ point along the two perpendicular axes, respectively, but
the magnitudes (or lengths) of the vectors $\boldsymbol{T}_{2n+1}$ and
$\boldsymbol{T}_{2n}$ are both geometrical series. This character allows us to
get the analytical expression of the infinite summation.\begin{figure}[ptb]
\centerline{\includegraphics[scale=1.5]{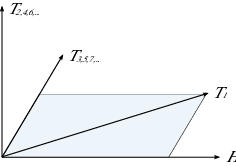}}\caption{Illustrating
figure showing how to calculate $T_{n}$ using the vector cross times (see
Eq.[4]).}%
\end{figure}

Summing up all the terms, we get an analytical expression for the position
operator of electron
\begin{align}
y_{H}(t)  &  =y(0)+\frac{it}{\hbar}\left(  {\epsilon}^{y}+T_{1}-\frac{T_{3}
}{|D|^{2}}\right) \nonumber\\
&  +\frac{T_{2}}{2|D|^{2}}\left[  \cos\left(  2|D|t/\hbar\right)  -1\right]
\nonumber\\
&  +\frac{iT_{3}}{2|D|^{3}}\left[  \sin\left(  2|D|t/\hbar\right)  \right]  ,
\end{align}
where $|D|=\sqrt{d_{1}^{2}+d_{2}^{2}+d_{3}^{2}}$.

\section{II. Analytical expression for electron trajectory: adiabatic
approximation}

The Hamiltonian for a two-dimensional topological insulator, a HgTe quantum
well with an inverted band structure described by the BHZ model, under a
uniform electric filed is
\[
H(k)_{\uparrow\downarrow}={H}_{0}(k)_{\uparrow\downarrow}+V(x_{i}),
\]
where ${H}_{0}(k)_{\uparrow\downarrow}=C-Dk^{2}\pm Ak_{x}\sigma^{x}
+Ak_{y}\sigma^{y}+(M-Bk^{2})\sigma^{z}$, $V(\mathbf{x})=-e\mathbf{E\cdot
r}\sigma^{z}$. After the proper unitary transformation $U(k)_{\uparrow
\downarrow}$ which can diagonalize the ${H}_{0}(k)$, i. e., $\tilde{H}
_{0}=U(k)_{\uparrow\downarrow}H_{0}(k)_{\uparrow\downarrow}U^{\dag
}(k)_{\uparrow\downarrow}$, then $H_{\uparrow\downarrow}^{\text{eff}}
=\tilde{H}_{0}(k)-e {\textstyle \sum\nolimits_{i=x,y}} E_{i}D_{i,\uparrow
\downarrow}$, where
\[
\tilde{H}_{0}(k)=C-Dk^{2}+\left(
\begin{array}
[c]{cc}%
-\sqrt{A^{2}k^{2}+(M-Bk^{2})^{2}} & 0\\
0 & \sqrt{A^{2}k^{2}+(M-Bk^{2})^{2}}%
\end{array}
\right)  ,
\]
\[
U(k)_{\uparrow\downarrow}=\left(
\begin{array}
[c]{cc}%
\frac{\mp\lbrack-Bk^{2}+M-\sqrt{A^{2}k^{2}+(M-Bk^{2})^{2}}]}{A(k_{x}\pm
ik_{y})\sqrt{1+\frac{(Bk^{2}-M+\sqrt{A^{2}k^{2}+(M-Bk^{2})^{2}})}{A^{2}k^{2}}
}} & \frac{-Bk^{2}+M+\sqrt{A^{2}k^{2}+(M-Bk^{2})^{2}}}{A(k_{x}\pm ik_{y}
)\sqrt{1+\frac{(Bk^{2}+M+\sqrt{A^{2}k^{2}+(M-Bk^{2})^{2}})}{A^{2}k^{2}}}}\\
\frac{1}{\sqrt{1+\frac{(Bk^{2}-M+\sqrt{A^{2}k^{2}+(M-Bk^{2})^{2}})}{A^{2}
k^{2}}}} & \frac{1}{\sqrt{1+\frac{(-Bk^{2}+M+\sqrt{A^{2}k^{2}+(M-Bk^{2})^{2}
})}{A^{2}k^{2}}}}%
\end{array}
\right)  ,
\]
and%
\[
\tilde{A}_{i}(k)_{\uparrow\downarrow}=-i\cdot U(k)_{\uparrow\downarrow
}\partial_{k_{i}}U^{\dag}(k)_{\uparrow\downarrow}\cdot U(k)_{\uparrow
\downarrow}\sigma^{z}U^{\dag}(k)_{\uparrow\downarrow}.
\]
\

The analytical expression is too\textit{ lengthy} to be omitted here. Adopting
the adiabatic approximation, we neglect the interband transitions, i.e.,
neglecting the off-diagonal matrix elements of $\tilde{A}$,
\[
F_{xy}(k)_{\uparrow\downarrow}=i[D_{i},D_{j}]=\left(
\begin{array}
[c]{cc}%
\pm\frac{A^{2}(M^{2}-B^{2}k^{4})}{2[A^{2}k^{2}+(M-Bk^{2})^{2}]^{2}} & 0\\
0 & \pm\frac{A^{2}(M^{2}-B^{2}k^{4})}{2[A^{2}k^{2}+(M-Bk^{2})^{2}]^{2}}%
\end{array}
\right)  .
\]

The equation of motion for the spin-up/down ($\uparrow\downarrow$)
negative/positive branch ($\lambda=\pm1$) can be written as:
\[
k_{i}=k_{i0}+\lambda\frac{eE_{i}}{\hbar}t,
\]%
\begin{align*}
\frac{dx_{\uparrow\downarrow}}{dt}  &  =-\frac{2D}{\hbar}k_{x}+\lambda
\frac{2B(M-Bk^{2})-A^{2}}{\hbar\sqrt{A^{2}k^{2}+(M-Bk^{2})^{2}}}k_{x}\pm
\frac{eE_{y}}{\hbar}\frac{A^{2}(M^{2}-B^{2}k^{4})}{2[A^{2}k^{2}+(M-Bk^{2}%
)^{2}]^{2}},\\
\frac{dy_{\uparrow\downarrow}}{dt}  &  =-\frac{2D}{\hbar}k_{y}+\lambda
\frac{2B(M-Bk^{2})-A^{2}}{\hbar\sqrt{A^{2}k^{2}+(M-Bk^{2})^{2}}}k_{y}\mp
\frac{eE_{x}}{\hbar}\frac{A^{2}(M^{2}-B^{2}k^{4})}{2[A^{2}k^{2}+(M-Bk^{2}%
)^{2}]^{2}}.
\end{align*}
When $E_{y}=k_{y}=0$, the integration of the topological term
{\cite{SCZhang01}}, $\mp\frac{eE_{x}}{\hbar}\frac{A^{2}(M^{2}-B^{2}k^{4}%
)}{2[A^{2}k^{2}+(M-Bk^{2})^{2}]^{2}}$, represents the electron's orbital
motion in the $y$ axis brought by the effective field strength $F_{xy}$,
\[
y_{\uparrow\downarrow}^{\text{orb}}=\int dk\frac{A^{2}(M^{2}-B^{2}k^{4}%
)}{2[A^{2}k^{2}+(M-Bk^{2})^{2}]^{2}}.
\]
To the second order of $k$,
\[
y_{\uparrow\downarrow}^{\text{orb}}=\mp\frac{A^{2}}{2M^{2}}k+O(k)^{3},
\]
and for the $k\rightarrow\infty$ limit,
\[
y_{\uparrow\downarrow}^{\text{orb}}\rightarrow\mp\frac{iA\pi\left(
\sqrt{\frac{B^{2}}{A^{2}-2BM-iA\sqrt{-A^{2}+4BM}}}-\sqrt{\frac{B^{2}}%
{A^{2}-2BM+iA\sqrt{-A^{2}+4BM}}}\right)  }{4\sqrt{-2A^{2}+8BM}},
\]
which is about $14.31$nm adopting the parameters in Ref. {\cite{MKonig}}. We
can see that this topological shift have a upper limit during a single
ballistic motion.

Next, we derive the analytical expression for the trembling motion, i.e., the
\textit{Zitterbewegung} {\cite{SCZhang02}. We assume the wave function has the
form of }$\left\vert \psi(x,t)\right\rangle =\exp{(-ie\mathbf{E}
\cdot\mathbf{r}\sigma}_{z}t/\hbar{)}\left\vert u(x,t)\right\rangle $, and
substitute $\left\vert \psi(x,t)\right\rangle $ into the Schr\"{o}dinger equation%

\[
i\hbar\partial_{t}\left\vert \psi(x,t)\right\rangle =[H_{0}(k)_{\uparrow
\downarrow}+{e\mathbf{E}\cdot\mathbf{r}\sigma}_{z}]\left\vert \psi
(x,t)\right\rangle ,
\]
where ${H}_{0}(k)_{\uparrow\downarrow}=C-Dk^{2}\pm Ak_{x}\sigma^{x}
+Ak_{y}\sigma^{y}+(M-Bk^{2})\sigma^{z}$, because
\begin{align*}
i\hbar\partial_{t}\left\vert \psi(x,t)\right\rangle  &  =[{e\mathbf{E}
\cdot\mathbf{r}\sigma}_{z}]\left\vert \psi(x,t)\right\rangle \\
+  &  \exp{(-ie\mathbf{E}\cdot\mathbf{r}\sigma}_{z}t/\hbar{)[}i\hbar
\partial_{t}\left\vert u(x,t)\right\rangle ],
\end{align*}
we get a time-dependent Schr\"{o}dinger equation
\[
i\hbar\partial_{t}\left\vert u(x,t)\right\rangle =H_{0}(k,t)\left\vert
u(x,t)\right\rangle ,
\]
where $k_{i}(t)=k_{i0}+\lambda\frac{eE_{i}}{\hbar}t$. We assume that
$\left\vert u(x,t)\right\rangle ={\textstyle \sum\nolimits_{\lambda}}
C_{\lambda}(t)\exp(-i\alpha_{\lambda})U^{\dag}(k)\left\vert \lambda
\right\rangle $, where $\left\vert \lambda\right\rangle $ represents any
eigenstate of $S_{z}$, that is $S_{z}\left\vert \lambda\right\rangle
=\lambda\left\vert \lambda\right\rangle $, so $U^{\dag}(k)_{\uparrow
\downarrow}\left\vert \lambda\right\rangle $ is the instant eigenstate of
$H_{0}(k)_{\uparrow\downarrow}$, $H_{0}(k)_{\uparrow\downarrow}U^{\dag
}(k)_{\uparrow\downarrow}\left\vert \lambda\right\rangle =\epsilon_{\lambda
}(t)U^{\dag}(k)_{\uparrow\downarrow}\left\vert \lambda\right\rangle $, where
$\epsilon_{\frac{1}{2}/\frac{3}{2}}(t)=C-Dk^{2}\mp\sqrt{A^{2}k^{2}
+(M-Bk^{2})^{2}}$. $\alpha_{\lambda}=(1/\hbar){\textstyle \int\nolimits_{0}
^{t}} \epsilon_{\lambda}(t^{\prime})dt^{\prime}$. Because%
\begin{align*}
i\hbar\partial_{t}\left\vert u(x,t)\right\rangle  &  =i\hbar{\textstyle \sum
\nolimits_{\lambda}} [\partial_{t}C_{\lambda}(t)]\exp(-i\alpha_{\lambda
})U^{\dag}(k)_{\uparrow\downarrow}\left\vert \lambda\right\rangle \\
&  + {\textstyle \sum\nolimits_{\lambda}} C_{\lambda}(t)[\epsilon_{\lambda
}(t)\exp(-i\alpha_{\lambda})]U^{\dag}(k)_{\uparrow\downarrow}\left\vert
\lambda\right\rangle \\
&  +i\hbar{\textstyle \sum\nolimits_{\lambda}} C_{\lambda}(t)\exp
(-i\alpha_{\lambda})[\partial_{t}U^{\dag}(k)_{\uparrow\downarrow}]\left\vert
\lambda\right\rangle ,
\end{align*}
and%
\[
H_{0}\left\vert u(x,t)\right\rangle ={\textstyle \sum\nolimits_{\lambda}}
C_{\lambda}(t)[\epsilon_{\lambda}(t)\exp(-i\alpha_{\lambda})]U^{\dag
}(k)_{\uparrow\downarrow}\left\vert \lambda\right\rangle
\]
thus
\[
{\textstyle \sum\nolimits_{\lambda}} [\partial_{t}C_{\lambda}(t)]\exp
(-i\alpha_{\lambda})U^{\dag}(k)_{\uparrow\downarrow}\left\vert \lambda
\right\rangle + {\textstyle \sum\nolimits_{\lambda}} C_{\lambda}%
(t)\exp(-i\alpha_{\lambda})[\partial_{t}U^{\dag}(k)_{\uparrow\downarrow
}]\left\vert \lambda\right\rangle =0.
\]
Act $U(k)_{\uparrow\downarrow}$ and $\left\langle \lambda^{\prime}\right\vert
$ to the left side,
\[
\lbrack\partial_{t}C_{\lambda^{\prime}}(t)]\exp(-i\alpha_{\lambda^{\prime}})+
{\textstyle \sum\nolimits_{\lambda}} C_{\lambda}(t)\exp(-i\alpha_{\lambda
})\left\langle \lambda^{\prime}\right\vert U(k)_{\uparrow\downarrow}%
\partial_{t}U^{\dag}(k)_{\uparrow\downarrow}\left\vert \lambda\right\rangle
=0.
\]

For the helicity state, $\left\vert 1/2\right\rangle =(1,0)^{T}$, $\left\vert
3/2\right\rangle =(0,1)^{T}$, we set the initial state $C_{1/2}(0)=1$,
$C_{3/2}(0)=0$. The adiabatic approximation assumes that $0\simeq
C_{3/2}(t)\ll C_{1/2}(t)$ is always satisfied, which leads to
\begin{align*}
\partial_{t}C_{1/2}(t)  &  \simeq-C_{1/2}(t)\left\langle 1/2\right\vert
U(k)_{\uparrow\downarrow}\partial_{t}U^{\dag}(k)_{\uparrow\downarrow
}\left\vert 1/2\right\rangle ,\\
\partial_{t}C_{3/2}(t)  &  \simeq-C_{1/2}(t)\exp(i\Delta\alpha)\left\langle
3/2\right\vert U(k)_{\uparrow\downarrow}\partial_{t}U^{\dag}(k)_{\uparrow
\downarrow}\left\vert 1/2\right\rangle ,\\
\Delta\alpha &  =\alpha_{3/2}-\alpha_{1/2}=(1/\hbar){\textstyle \int
\nolimits_{0}^{t}} \Delta\epsilon(t^{\prime})dt^{\prime}.
\end{align*}

Because $y_{\uparrow\downarrow}^{\text{ZB}}=C_{3/2}^{\ast}C_{1/2}
e^{-i\Delta\alpha}\left\langle 3/2\right\vert iU(k)_{\uparrow\downarrow
}\partial_{k_{y}}U^{\dag}(k)_{\uparrow\downarrow}\left\vert 1/2\right\rangle
+$h$.$c$.$, Then we can calculate the $U(k)_{\uparrow\downarrow}\partial
_{t}U^{\dag}(k)_{\uparrow\downarrow}$ and $U(k)_{\uparrow\downarrow}
\partial_{k_{y}}U^{\dag}(k)_{\uparrow\downarrow}$. When $E_{y}=\partial
k_{y}/\partial t=0$,
\[
U(k)_{\uparrow\downarrow}\partial_{t}U^{\dag}(k)_{\uparrow\downarrow
}=U(k)_{\uparrow\downarrow}\frac{\partial}{\partial k_{x}}U^{\dag
}(k)_{\uparrow\downarrow}\cdot\frac{\partial k_{x}}{\partial t},
\]
at small $k$,
\begin{align*}
\left.  U(k)_{\uparrow\downarrow}\frac{\partial}{\partial k_{x}}U^{\dag
}(k)_{\uparrow\downarrow}\right\vert _{k=0}  &  =\left[
\begin{array}
[c]{cc}%
0 & \pm\frac{A}{2M}\\
\mp\frac{A}{2M} & 0
\end{array}
\right]  +O(k^{2}),\\
\left.  U(k)_{\uparrow\downarrow}\frac{\partial}{\partial k_{y}}U^{\dag
}(k)_{\uparrow\downarrow}\right\vert _{k=0}  &  =\left[
\begin{array}
[c]{cc}%
0 & -\frac{iA}{2M}\\
-\frac{iA}{2M} & 0
\end{array}
\right]  +O(k^{2}),
\end{align*}

Notice that the $U(k)_{\uparrow\downarrow}\partial_{k_{y}}U^{\dag
}(k)_{\uparrow\downarrow}$ and $\left\langle 3/2\right\vert iU(k)_{\uparrow
\downarrow}\partial_{k_{y}}U^{\dag}(k)_{\uparrow\downarrow}\left\vert
1/2\right\rangle $ is only finite for the diagonal and off-diagonal matrix
elements, respectively. Therefore
\begin{align*}
\partial_{t}C_{1/2}(t)_{\uparrow\downarrow}  &  =0,\\
\partial_{t}C_{3/2}(t)_{\uparrow\downarrow}  &  =\mp\frac{A}{2M}\frac{eE_{x}
}{\hbar}\exp(i\Delta\alpha).
\end{align*}

Adopting the adiabatic approximation, $\epsilon_{\lambda}(t)$ are slowly
varying functions of the time $t$,
\[
C_{3/2}(t)_{\uparrow\downarrow}=\mp\frac{A}{2M}\frac{eE_{x}}{\Delta
\epsilon(t)}\exp(i\Delta\alpha).
\]

Finally we get
\[
y_{\uparrow\downarrow}^{\text{ZB}}=\pm\frac{A^{2}}{2M^{2}}\frac{eE_{x}}%
{\Delta\epsilon(t)}\sin(\frac{\Delta\epsilon(t)}{\hbar}t),
\]
where $\Delta\epsilon(t)=2\sqrt{A^{2}k^{2}+(M-Bk^{2})^{2}}=2\left\vert
M\right\vert +O(k^{2})$. At small $k$,
\[
y_{\uparrow\downarrow}^{\text{ZB}}\simeq\pm\frac{A^{2}}{4M^{2}}\frac{eE_{x}%
}{\left\vert M\right\vert }\sin(\frac{\Delta\epsilon(t)}{\hbar}t).
\]

When we consider the trajectory of an initially resting electron/hole in 2D TI
under an AC electric field in the $x$-direction $E_{x}=E_{0}\cos(\omega
t/\hbar)$, the equation of orbital motion for the spin-up/down ($\uparrow
\downarrow$) electron/hole ($\lambda=\pm1$) is%
\[
k_{x}=\lambda\frac{eE_{0}}{\hbar}\frac{\hbar}{\omega}\sin(\omega t/\hbar),
\]%
\begin{align*}
\frac{dx_{\uparrow\downarrow}}{dt} &  =-\frac{2D}{\hbar}k_{x}+\lambda
\frac{2B(M-Bk^{2})-A^{2}}{\hbar\sqrt{A^{2}k^{2}+(M-Bk^{2})^{2}}}k_{x},\\
\frac{dy_{\uparrow\downarrow}}{dt} &  =\mp\frac{eE_{x}}{\hbar}\frac
{A^{2}(M^{2}-B^{2}k^{4})}{2[A^{2}k^{2}+(M-Bk^{2})^{2}]^{2}}.
\end{align*}
To the second order of $k$,%
\begin{align*}
\frac{dx_{\uparrow\downarrow}}{dt} &  =-\frac{2D}{\hbar}k_{x}+\lambda
\frac{2BM-A^{2}}{\hbar\left\vert M\right\vert }k_{x},\\
\frac{dy_{\uparrow\downarrow}}{dt} &  =\mp\frac{eE_{x}}{\hbar}\frac{A^{2}%
}{2M^{2}},
\end{align*}
then we can get the orbital motion under the adiabatic approximation
\begin{align*}
x_{\uparrow\downarrow}^{\text{orb}} &  =\frac{eE_{0}}{\hbar^{2}}\left(
\lambda\frac{A^{2}-2BM}{\left\vert M\right\vert }+2D\right)  \left(
\frac{\hbar}{\omega}\right)  ^{2}\left[  \cos(\omega t/\hbar)-1\right]  ,\\
y_{\uparrow\downarrow}^{\text{orb}} &  =\mp\frac{A^{2}}{2M^{2}}\frac{eE_{0}%
}{\hbar}\frac{\hbar}{\omega}\sin(\omega t/\hbar).
\end{align*}

\section{III. The numerical calculation of electron trajectory $y(t)$}

We assume that the electric field $\emph{E}=(-E,0,0)$ points along the $-x$
axis without loss of any generality, thus
\[
H_{\uparrow\downarrow}(k)={C-Dk}^{2}+(\pm Ak_{x},Ak_{y},M-Bk^{2})\cdot
\sigma-eEx\cdot\sigma_{z},
\]
or
\[
H(k)={C-Dk}^{2}+Ak_{x}\cdot s_{1}+Ak_{y}\cdot s_{2}+(M-Bk^{2})\cdot
s_{3}-eEx\cdot s_{3},
\]
where we define
\[
s_{1}=\left[
\begin{array}
[c]{cc}%
\sigma_{x} & 0\\
0 & -\sigma_{x}%
\end{array}
\right]  ,s_{2}=\left[
\begin{array}
[c]{cc}%
\sigma_{y} & 0\\
0 & \sigma_{y}%
\end{array}
\right]  ,s_{3}=\left[
\begin{array}
[c]{cc}%
\sigma_{z} & 0\\
0 & \sigma_{z}%
\end{array}
\right]  ,
\]
\[
s_{4}=\left[
\begin{array}
[c]{cc}%
\sigma_{x} & 0\\
0 & \sigma_{x}%
\end{array}
\right]  ,s_{5}=\left[
\begin{array}
[c]{cc}%
\sigma_{y} & 0\\
0 & -\sigma_{y}%
\end{array}
\right]  ,s_{6}=\left[
\begin{array}
[c]{cc}%
\sigma_{z} & 0\\
0 & -\sigma_{z}%
\end{array}
\right]  ,
\]
then
\begin{align*}
\frac{d\left\langle k_{x}\right\rangle }{dt}  &  =\frac{eE}{\hbar}\left\langle
s_{3}\right\rangle ,\\
\frac{d\left\langle k_{y}\right\rangle }{dt}  &  =0,
\end{align*}
\begin{align*}
\frac{d\left\langle x\right\rangle }{dt}  &  =-\frac{2D}{\hbar}\left\langle
k_{x}\right\rangle +\frac{A}{\hbar}\left\langle s_{1}\right\rangle -\frac
{2B}{\hbar}\left\langle k_{x}\right\rangle \left\langle s_{3}\right\rangle ,\\
\frac{d\left\langle y\right\rangle }{dt}  &  =-\frac{2D}{\hbar}\left\langle
k_{y}\right\rangle +\frac{A}{\hbar}\left\langle s_{2}\right\rangle -\frac
{2B}{\hbar}\left\langle k_{y}\right\rangle \left\langle s_{3}\right\rangle ,
\end{align*}
\begin{align*}
\frac{d\left\langle s_{1}\right\rangle }{dt}  &  =-2\frac{M-B\left\langle
k\right\rangle ^{2}-eE\left\langle x\right\rangle }{\hbar}\left\langle
s_{5}\right\rangle +\frac{2A}{\hbar}\left\langle k_{y}\right\rangle
\left\langle s_{6}\right\rangle ,\\
\frac{d\left\langle s_{2}\right\rangle }{dt}  &  =2\frac{M-B\left\langle
k\right\rangle ^{2}-eE\left\langle x\right\rangle }{\hbar}\left\langle
s_{4}\right\rangle -\frac{2A}{\hbar}\left\langle k_{x}\right\rangle
\left\langle s_{6}\right\rangle ,\\
\frac{d\left\langle s_{3}\right\rangle }{dt}  &  =-\frac{2A}{\hbar
}\left\langle k_{y}\right\rangle \left\langle s_{4}\right\rangle +\frac
{2A}{\hbar}\left\langle k_{x}\right\rangle \left\langle s_{5}\right\rangle ,
\end{align*}
\begin{align*}
\frac{d\left\langle s_{4}\right\rangle }{dt}  &  =-2\frac{M-B\left\langle
k\right\rangle ^{2}-eE\left\langle x\right\rangle }{\hbar}\left\langle
s_{2}\right\rangle +\frac{2A}{\hbar}\left\langle k_{y}\right\rangle
\left\langle s_{3}\right\rangle ,\\
\frac{d\left\langle s_{5}\right\rangle }{dt}  &  =2\frac{M-B\left\langle
k\right\rangle ^{2}-eE\left\langle x\right\rangle }{\hbar}\left\langle
s_{1}\right\rangle -\frac{2A}{\hbar}\left\langle k_{x}\right\rangle
\left\langle s_{3}\right\rangle ,\\
\frac{d\left\langle s_{6}\right\rangle }{dt}  &  =-\frac{2A}{\hbar
}\left\langle k_{y}\right\rangle \left\langle s_{1}\right\rangle +\frac
{2A}{\hbar}\left\langle k_{x}\right\rangle \left\langle s_{2}\right\rangle .
\end{align*}
For a given initial state, we can calculate the electron trajectory
$\left\langle y(t)\right\rangle $ involving with the time $t$ numerically.